\documentclass[sigconf]{acmart}
\renewcommand\footnotetextcopyrightpermission[1]{} 
\usepackage{booktabs} 

\usepackage{blindtext}
\usepackage{amsmath,amssymb}
\usepackage{mathtools}   
\usepackage[linesnumbered, ruled]{algorithm2e} 
\usepackage[british,UKenglish]{babel}
\usepackage{array}
\usepackage{booktabs}

\DeclareMathOperator*{\argmin}{arg\,min}
\DeclareMathOperator*{\argmax}{arg\,max}

\newcommand{\premise}[1]{\textcolor{black}{#1}}  
\newcommand{\evokepr}[1]{\textcolor{black}{#1}}  
\newcommand{\improve}[1]{}                       
\newcommand{\improvex}[2]{\textcolor{black}{#2}} 
\newcommand{\fix}[2]{\textcolor{black}{#2}}      

\newcommand{\theme}[1]{}
\newcommand{\rheme}[1]{\textcolor{black}{#1}}

\setcopyright{acmcopyright}

\begin{document}
\title{Limits to Surprise in Recommender Systems}


\author{Andre Paulino de Lima}
\orcid{0000-0002-3148-6686}
\affiliation{%
  \institution{Universidade de Sao Paulo}
}
\email{andre.p.lima@usp.br}

\author{Sarajane Marques Peres}
\orcid{0000-0003-3551-6480}
\affiliation{%
	\institution{Universidade de Sao Paulo}
}
\email{sarajane@usp.br}


\begin{abstract}

\theme{goal}
In this study, we address the challenge of measuring the ability of a recommender system to make surprising recommendations.
Although current evaluation methods make it possible to determine if two algorithms can make recommendations with a significant difference in their average surprise measure, it could be of interest to our community to know how competent an algorithm is at embedding surprise in its recommendations, without having to resort to making a direct comparison with another algorithm.
We argue that a) \rheme{surprise is a finite resource in a recommender system, b) there is a limit to how much surprise any algorithm can embed in a recommendation, and c) this limit can provide us with a scale against which the performance of any algorithm can be measured}.
\theme{method}
By exploring these ideas, it is possible to define the concepts of maximum and minimum potential surprise and design a surprise metric called ``normalised surprise'' that employs these limits to potential surprise.
Two experiments were conducted to test the proposed metric.
The aim of the first was to validate the quality of the estimates of minimum and maximum potential surprise produced by a greedy algorithm.
The purpose of the second experiment was to analyse the behaviour of the proposed metric using the MovieLens dataset.
\theme{results, commentary}
The results confirmed the behaviour that was expected, and showed that the proposed surprise metric is both effective and consistent for differing choices of recommendation algorithms, data representations and distance functions.

\end{abstract}

%
%


\vspace{5em}


\maketitle

\newpage
\section{Introduction}
\improve{from Jim Hesson: approach literature review, report on the importance of the research, and elaborate on the gap in the literature that the work fills.}
\theme{move 1 - begin by defining the larger general territory or context from which the topic of your study develops}
It has been well established that the effect of beyond-accuracy properties on \fix{check if this expression (user overall satisfaction) is used in community jargon/ I think I read it in Ekstrand, the work on user perception of changes in recommenders}{user satisfaction} is a critical success factor in deploying a recommender system \cite{mcnee2006being, herlocker2004evaluating}.
Among these properties, surprise has recently been the subject of several studies owing to its links to serendipity \cite{adamopoulos2011unexpectedness, kaminskas2014measuring, silveira2017framework} and the problem of over-specialisation in content-based recommender systems \cite{HandbookRS2015Cap4}, as well as its importance in some application domains \cite{mourao2017surprises}.

In the literature, the notion of surprise generally reflects the capacity to make recommendations that are dissimilar from the items known to a given user: the more a recommended item is dissimilar, the more it is surprising \cite{kaminskas2014measuring,adamopoulos2011unexpectedness,zhang2012auralist}.
\theme{move 2 - point out a gap or lack of knowledge that exists in the literature related to the topic of your study}
Current surprise metrics and evaluation methods allow us to estimate the average surprise in recommendations produced by an algorithm.
Given a set of algorithms and a fixed experimental setting, statistical tools can be used to estimate if there is a significant difference in the average degree of surprise between recommendations produced by any two algorithms.
Although this kind of approach can be successfully applied when selecting which algorithm might be more promising for an application domain, it is unable to provide a common measurement scale that remains consistent in different experimental settings; nor does it reveal how much room there is for improvement with respect to surprise.

\theme{move 3 - indicate how your study fills this gap}
In this study, we express the view that surprise can be regarded as a system resource: at any given time, a recommender system has a limited ``stock of surprise'' that is available to each user.
This theoretical ``stock of surprise'' is referred to here as ``maximum potential surprise''.
The recommendation algorithm, which is designed to optimise a set of objectives, controls how much surprise is embedded in each recommendation it produces.
However, there is a limit to how much of the available surprise any recommendation algorithm can embed in a recommendation.
By pursuing this line of thought, we were able to devise a surprise metric, called ``normalised surprise'', which provides a measurement scale in which the meaning remains consistent across different settings.
For example, it can provide the information that, on average, a recommender system has embedded 20\% of the available surprise in each recommendation it produces.
As a result, this means that there is still 80\% of the available surprise that can be appropriated by the system.

\theme{provide a roadmap}
The remainder of this paper is structured as follows: Section \ref{sec:relatedwork} examines related work on surprise evaluation; in Section \ref{sec:sp_intro} a theoretical model is designed for potential surprise and the proposed surprise metric; Section \ref{sec:experiments} describes the experiments conducted to validate the proposed metric through both an ancillary synthetic dataset and the popular MovieLens dataset, and the results are discussed in Section \ref{sec:discussion}.
Finally, we conclude by summarising the work and making suggestions for possible future work in Section \ref{sec:conclusion}.

\section{Related Work}
\label{sec:relatedwork}
Before providing a review of relevant work related to this research, in this section there is an examination of the properties of a recommender system that are related to surprise (Section \ref{sec:surprise_property}).
This is followed by a discussion of several surprise-related metrics that have been proposed in the literature (Section \ref{sec:surprise_metrics}) and the one plus random off-line evaluation method for surprise (Section \ref{sec:surprise_evaluation}).

\subsection{The Surprise Property}
\label{sec:surprise_property}
\theme{Tell a story about the field, focusing on concepts that you are working with. Try to organise the field in such a way as you can see how the work evolves from start to finish.}
\improvex{LLA: comma after subject}{The challenge of discovering new items that might be useful to a user has been focus of a large number of works in literature on recommender systems.}
In general, the approach involves finding new items that bear some similarity to items which have been given good ratings by some users.
An even greater challenge is to find new items that do not resemble items known to a user, yet would still be useful to them.
This would be a serendipitous recommendation.

\citeauthor{herlocker2002empirical} \cite{herlocker2002empirical} offers a definition of serendipity that is usually cited by work in this area: \textit{``A serendipitous recommendation helps the user find a surprisingly interesting item he might not have otherwise discovered.''}
In a sense, this definition supports a perspective whereby serendipity, as a system property, results from the interaction of two other and more fundamental properties: surprise and relevance.
Being surprising and relevant (or useful) to a user are the basic requirements of a serendipitous recommendation.

It has been recently pointed out by \citeauthor{kaminskas2016diversity} \cite{kaminskas2016diversity} that there is a conceptual overlap between the properties of novelty or unexpectedness and the notion of surprise.
In this study, we subscribe to the categorisation suggested by these authors, in which a) novelty is related to the notion of an item being popular, and thus is not directly related to serendipity, b) unexpectedness usually conveys the same notion as surprise, and, as mentioned earlier, c) surprise can be regarded as a component of serendipity.

In view of this, our focus is on the metrics employed for estimating surprise, serendipity and unexpectedness.
This review sets out by pointing out that several authors have approached the problem of measuring these properties by adopting strategies that, although clearly  distinct from each other, have some key features in common.
Each strategy is analysed on the basis of three factors:

\textit{Intrinsic vs extrinsic evaluation}\footnote{Here, intrinsic vs extrinsic evaluation is an analogy to the same dichotomy employed in clustering quality evaluation methods \cite{han2011data}.}: some studies have defined metrics that only use data that are within the system under evaluation \cite{akiyama2010proposal, zhang2012auralist, kaminskas2014measuring}, while others have defined metrics that use data made available by an external system (often referred to as PPM - Primitive Prediction Model) in addition to the internal data \cite{murakami2007metrics, ge2010beyond, adamopoulos2011unexpectedness}

\textit{Subjective vs objective view}: some metrics assume that surprise is subjective in nature, since it depends on the set of items known to each user \cite{murakami2007metrics, adamopoulos2011unexpectedness, zhang2012auralist, kaminskas2014measuring}, while others view surprise as a property of the item itself \cite{ge2010beyond, akiyama2010proposal} and, thus, is independent of the users.

\textit{Reductionist vs non-reductionist approach}: some authors have employed a reductionist approach, in so far as they seek to isolate the surprise and relevance components of serendipity and examine them as separate metrics \cite{murakami2007metrics, akiyama2010proposal, adamopoulos2011unexpectedness, zhang2012auralist, kaminskas2014measuring}, while others have proposed metrics that treat surprise and relevance in a more integrated way \cite{ge2010beyond}.

\subsection{Surprise Metrics}
\label{sec:surprise_metrics}
\begin{figure}[H]
	\includegraphics[scale=.25]{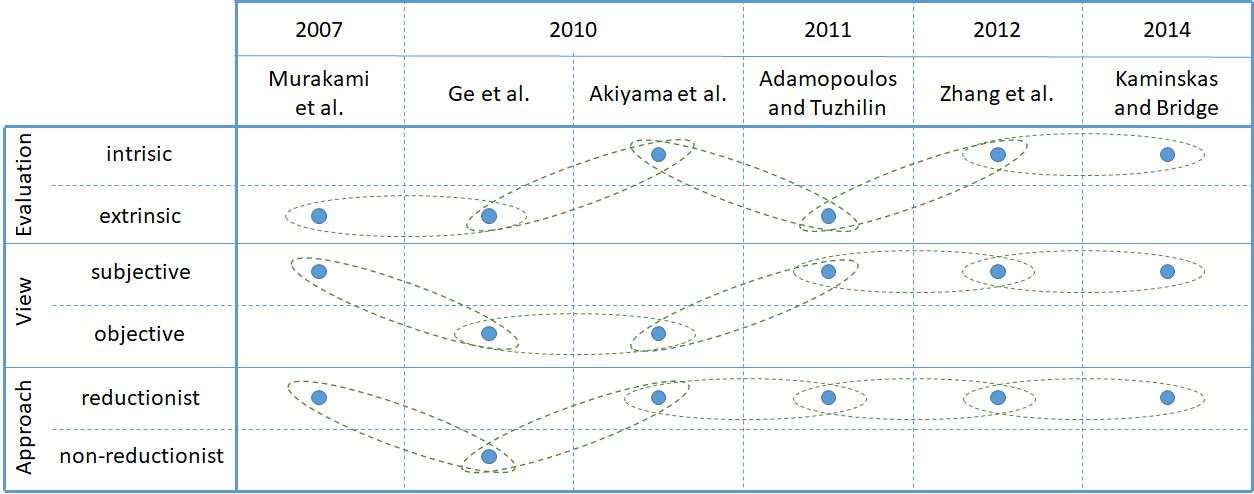}
	\caption{Evolution of surprise-related metrics over time.}
	\label{fig:trends}
\end{figure}

In this section, there is a review of six surprise-related metrics in the literature.
Figure \ref{fig:trends} provides a summary of the review, and illustrates how they are positioned with regard to the factors described in Section \ref{sec:surprise_property}.
As can be seen in the diagram, the metrics include the year of publication, and the ellipses show trends or changes in the factors.
This review is not meant to be exhaustive but rather aims to capture the approaches that have evolved over a period of time. 

\subsubsection{\textbf{A metric for unexpectedness}}

\citeauthor{murakami2007metrics} \cite{murakami2007metrics} have proposed a metric to evaluate serendipity that explores the idea that a serendipitous recommendation must be ``non-obvious'', whereas recommendations made by a PPM are expected to be obvious.
As shown in Equation \ref{eq:murakami_unexp}, the metric is calculated from a recommendation list $L$ produced for the user $u$ by the system under evaluation. 
In this definition, the predicate $rscore$ accounts for the predicted relevance of an item $L_i$ to the user $u$, while $isrel$ accounts for surprise, and reflects the degree to which an item $L_i$ is similar to items rated highly by the user (i.e. based on a subjective view).
Since there are separate metrics for surprise and relevance, it can be assumed that a reductionist approach is being adopted.
Note that the predicate $rscore$, in Equation \ref{eq:murakami_score}, includes the relevance predicted by both the system under evaluation ($Pr$) and the external system ($Pr^*$), and can thus be regarded as an extrinsic evaluation.
\begin{equation}
\label{eq:murakami_unexp}
unexp(L, u) =
\frac{1}{|L|} \sum_{i=1}^{|L|} rscore(L_i, u) \times isrel(L_i, u)
\end{equation}
\begin{equation}
\label{eq:murakami_score}
rscore(L_i, u) = max(Pr(L_i, u) - Pr^*(L_i, u), 0).
\end{equation}

\subsubsection{\textbf{A metric for serendipity}}
\citeauthor{ge2010beyond} \cite{ge2010beyond} devised a metric for evaluating serendipity that follows the same line of thought pursued by \citeauthor{murakami2007metrics} \cite{murakami2007metrics}, although the external system is employed in a different way.
As shown in Equation \ref{eq:ge_srdp}, $srdp$ is applied to a recommendation list $L^\delta$, and estimates the \textit{usefulness} of each item $L^\delta_i$, that accounts for relevance.
In Equation \ref{eq:ge_unexp}, $L^\delta$ is defined as a list that consists of the elements recommended to the user $u$ by the system under evaluation ($L$) and that do not appear in the list drawn up for user $u$ by an external system ($L^*$).
This means that $L^\delta$ comprises non-obvious, unexpected items, and hence only accounts for surprise.
\begin{equation}
\label{eq:ge_srdp}
srdp(L^\delta, u) =
\frac{1}{|L^\delta|} \sum_{i=1}^{|L^\delta|} \mbox{\textit{usefulness}}(L^\delta_i, u)
\end{equation}
\begin{equation}
\label{eq:ge_unexp}
L^\delta = L \backslash L^*
\end{equation}
In addition, as there is no specific metric for surprise in Equations \ref{eq:ge_srdp} and \ref{eq:ge_unexp}, it can be assumed that $srdp$ adopts a non-reductionist approach.
Note that $srdp$ operates in an objective way, since estimating surprise ($L^\delta$) does not involve evaluating the degree to which new items are similar to items already known to the user.

\subsubsection{\textbf{A metric for general unexpectedness}}
\citeauthor{akiyama2010proposal} \cite{akiyama2010proposal} set out a metric called ``general unexpectedness'' that explores a combinatorial intuition: an item that shows a rare combination of attributes must be treated as unexpected.
It assumes that each item has some content combined with it, and that such a content can be described by a set of attributes. 
This usually is the case with content-based recommenders \cite{HandbookRS2015Cap4}.
As shown in Equation \ref{eq:akiyama_unexp}, the \textit{unexp} metric is estimated for $L$, the recommendation list produced to user $u$ by the system under evaluation, and this aggregates the \textit{uscore} obtained for each item $L_i$.
The \textit{uscore}, defined in Equation \ref{eq:akiyama_score}, is the reciprocal of the \fix{issue: this is not an average}{average joint probability} estimated for each pair of attributes of $L_i$. 
In this equation, $A(L_i)$ represents the set of attributes that describe $L_i$, $N_a$ denotes the number of items in the repository that have attribute $a$ and $N_{a, b}$ is the number of items that have both attributes $a$ and $b$.
Thus an objective view is adopted since surprise can be seen as a property of the content of an item.
\improvex{LLA: metrics previously described or previously described metrics}{Unlike the metrics previously described}, this metric does not employ an external system (i.e. an intrinsic evaluation). 
In addition, it should be noted that this metric only accounts for surprise, and it thus adopts a reductionist approach.
\begin{equation}
\label{eq:akiyama_unexp}
unexp(L)= \frac{1}{|L|}\sum_{i = 1}^{|L|} uscore(L_i)
\end{equation}
\begin{equation}
\label{eq:akiyama_score}
uscore(L_i)= \bigg[\frac{1}{|A(L_i)|} \sum_{a,b \in A(L_i)} \frac{N_{a,b}}{N_a + N_b - N_{a, b}} \bigg]^{-1}
\end{equation}

\subsubsection{\textbf{A metric for unexpectedness}}
\citeauthor{adamopoulos2011unexpectedness}  \cite{adamopoulos2011unexpectedness} propose a metric for unexpectedness that examines an intuition about user expectation: an item is expected for user $u$ if it is known to them or bears some similarity to items known to them.
As shown in Equation \ref{eq:adamo_unexp}, the \textit{unexp} metric is calculated from L, the recommendation list produced for user $u$ by the system under evaluation, and $L^s$, a list of obvious, expected items that is defined in Equation \ref{eq:adamo_helper}. 
In that equation, $L^*$ is a recommendation list produced for user $u$ by an external system, $E_u$ represents the set of items that have been rated by user $u$, and \improvex{LLA: is the last phrase modifying the term "similar"?}{the predicate \textit{neighbours} represents the set of items in the repository ($I$) that are similar to the items in $E_u$ up to some degree specified by threshold parameters in $\theta$}.
This approach adopts an external system (extrinsic evaluation), the metric only accounts for surprise (through a reductionist approach), and adheres to a subjective view, since it takes account of past experience of the user.
\begin{equation}
\label{eq:adamo_unexp}
unexp(L, L^s) = \frac{1}{|L|}|L \backslash L^s|
\end{equation}
\begin{equation}
\label{eq:adamo_helper}
L^s(u) = L^* \cup E_u \cup neighbours(I, E_u, \theta)
\end{equation}

\subsubsection{\textbf{The unserendipity metric}}
\citeauthor{zhang2012auralist} \cite{zhang2012auralist} explore the idea that a serendipitous recommendation must be dissimilar to items known to the user, in a semantic sense. 
It resembles the metric proposed by \citeauthor{akiyama2010proposal} \cite{akiyama2010proposal}, since it assumes that each item is combined with some content, but in this case content attributes are represented as vectors in $\mathbb{R}^m$ instead of sets.
As is shown in Equation \ref{eq:zhang_unsrdp}, the metric is computed from the recommendation list drawn up for user $u$ by the system under evaluation ($L$), and results in a score that is the average cosine similarity obtained from the items in $L$ and the set of items known to the user ($E_u$).
a) This approach does not employ an external system (intrinsic evaluation); b) the metric only accounts for surprise (reductionist approach) and c) it adheres to a subjective view of surprise.
\improvex{LLA}{It should be noticed that, unlike the metrics shown earlier, $unsrdp$ is scale-inverted, since the lower the score, the more surprising the $L$ is}.
\begin{equation}
\label{eq:zhang_unsrdp}
unsrdp(L, u) = \frac{1}{|L||E_u|} \sum_{i \in L}\sum_{j \in E_u} cossim(i, j)
\end{equation}

\subsubsection{\textbf{A metric for surprise}}
In a similar way to \citeauthor{zhang2012auralist} \cite{zhang2012auralist}, \citeauthor{kaminskas2014measuring} \cite{kaminskas2014measuring} argue that a surprising recommendation must be dissimilar to items known to the user, but does not require that this dissimilarity should be semantic in nature.
They also explore the interplay between the notions of distance and similarity\footnote{Given a metric for distance, a similarity metric can be derived, and vice-versa \cite{deza2009encyclopedia}.}.
Equation \ref{eq:kam_surprise} shows that the metric is calculated from the recommendation list produced for user $u$ by the system under evaluation ($L$), and produces the average surprise computed for each item in $L$.
The surprise of an item $i$ in $L$ is estimated as either a) the minimum distance between $i$ and each item known to the user ($E_u$), as described in Equation \ref{eq:kam_sidist}, or b) the maximum degree of similarity between the same items, as shown in Equation \ref{eq:kam_sisim}.
The predicate $dist$ is defined as the Jaccard distance between the set of attributes recovered from contents linked to items $i$ and $j$, while the predicate $sim$ computes the normalised pointwise mutual information score (NPMI) \cite{bouma2009normalized} for the same items.
This approach does not employ an external system (intrinsic evaluation); the metric only accounts for surprise (reductionist approach) and supports a subjective view of surprise, since it takes account of the past experience of the users.
\begin{equation}
\label{eq:kam_surprise}
surprise(L, u) = \frac{1}{|L|} \sum_{i \in L} S_i(i, E_u)
\end{equation}
\begin{equation}
\label{eq:kam_sidist}
S_i(i, E_u) = \min\limits_{j \in E_u} dist(i, j)
\end{equation}
\begin{equation}
\label{eq:kam_sisim}
S_i(i, E_u) = \max\limits_{j \in E_u} sim(i, j)
\end{equation}

\theme{Show how the body of work holds together in some philosophical or technological way}
In summary, we argue that all metrics described involve (in an abstract sense) a notion of distance in their surprise component, when it is applied to a) known and unknown items (subjective view) \cite{murakami2007metrics, adamopoulos2011unexpectedness, zhang2012auralist, kaminskas2014measuring}, b) expected and unknown items (extrinsic evaluation) \cite{murakami2007metrics, ge2010beyond, adamopoulos2011unexpectedness}, or c) the content linked to different items \cite{akiyama2010proposal, adamopoulos2011unexpectedness, zhang2012auralist, kaminskas2014measuring}:
\begin{itemize}
	\item \citeauthor{murakami2007metrics} \cite{murakami2007metrics}: the predicate \textit{isrel} measures how similar an item is to the items known to a user.
	\item \citeauthor{ge2010beyond} \cite{ge2010beyond}: $L^\delta$ is the difference between sets of items, thus it is a distance based on a combinatorial intuition;
	\item \citeauthor{akiyama2010proposal} \cite{akiyama2010proposal}: the predicate \textit{uscore} is the reciprocal of a measure of similarity given by joint-probability;
	\item \citeauthor{adamopoulos2011unexpectedness}  \cite{adamopoulos2011unexpectedness}: the predicate $unexp$ is the difference between sets;
	\item \citeauthor{zhang2012auralist} \cite{zhang2012auralist}: the predicate $unsrdp$ is defined as the reciprocal of geometric similarity, and is thus a distance;
	\item \citeauthor{kaminskas2014measuring} \cite{kaminskas2014measuring}: the predicate $S_i$ directly specifies the distance and similarity functions.
\end{itemize}

\theme{End up with the conclusion that the present work is needed.}
Finally, if one accepts the idea that surprise is a form of distance, and there is a tendency to separate surprise from relevance when seeking serendipity, it can be argued that it might be fruitful to tackle the problem of evaluating surprise from an \fix{is there a better term?}{informational} perspective, by trying to answer the following questions:
\begin{itemize}
	\item Are there limits to how much surprise a recommender can offer to a user? 
	\item Are there limits to how surprising a recommendation list can be?
	\item \improvex{LLA: scale up in which?}{If these limits exist, is it possible to use them to create a scale up in which the performance of a system can be measured?}
	\item \improvex{LLA: comma between subject and predicate}{If these limits exist, do decisions on how to represent data or which distance function to employ influence them?}
\end{itemize}

The current metrics for surprise do not address these questions, and this study seeks to fill in the gaps.

\subsection{Evaluation of Surprise}
\label{sec:surprise_evaluation}
All the metrics described in Section \ref{sec:surprise_metrics} evaluate the surprise\footnote{In this subsection, the term surprise will also encompass both serendipity and unexpectedness predicates described in section \ref{sec:surprise_metrics}.} of a recommendation list that is produced by the system that is being evaluated. 
An evaluation method is required to obtain an estimate of how the system performs with regard to surprise. 
Most studies follow a statistical procedure to compute this kind of estimate: a sample of users is selected, recommendation lists are produced, surprise evaluations are made, and the average is calculated.

\theme{One-plus-random}
On the other hand, in response to a general consensus about the limited ability of the accuracy metrics to evaluate the performance of recommenders in top-N recommendation tasks, a new off-line evaluation method called ``one plus random'' was designed to estimate the recall of a recommender system \cite{cremonesi2008evaluation, bellogin2011precision}.
This method follows the intuition that, in a sufficiently large set $L_1$ that consists of items unknown to user $u$, most of these items are irrelevant to $u$.
Supposing that an item that is highly rated by $u$, namely $i^*$, is added to $L_1$, if an algorithm attributes to $i^*$ a score such that $i^*$ ranks among the top-N items $L_1$, then the algorithm has succeeded in the task.
In a recent study, \citeauthor{kaminskas2014measuring} \cite{kaminskas2014measuring} adapted this method to estimate the degree of surprise of a recommender system.
The intuition behind the one plus random method is retained and, in addition to computing an estimate for recall, it also computes the average surprise obtained from the recommendation lists produced for a sample of users.

\section{A theoretical model for surprise}
\label{sec:sp_intro}
Before addressing the questions posed in Section \ref{sec:surprise_metrics}, it should be made clear what properties a metric for surprise should have.

\premise{\textit{Surprise must be subjective}}: \citeauthor{barto2013novelty} \cite{barto2013novelty} carried out a review of the concepts of surprise and novelty in cognitive science, as well as their quantitative models\footnote{The notion of surprise in \citeauthor{kaminskas2016diversity} \cite{kaminskas2016diversity} is closer to the notion of novelty in cognitive science than that of surprise, as described in \citeauthor{barto2013novelty} \cite{barto2013novelty}.}.
\citeauthor{reisenzein2017review} \cite{reisenzein2017review} examined to which extent the experimental evidence supports different quantitative models of surprise.
Both these studies portray these phenomena as subjective in nature, and show that their quantitative models usually involve some form of subjective probability, which may represent expectations or beliefs held by an individual.
From a more intuitive standpoint, and when applied to recommender systems, this idea can be illustrated through the Scenario 1:

\begin{itemize}
	\item [a)] Suppose items $i$ and $j$ are similar to each other;
	\item [b)] Suppose user $u_1$ has not been exposed to items $i$ and $j$, and all the items known to $u_1$ are very dissimilar from $i$ and $j$;
	\item [b)] Suppose user $u_2$ has been exposed to item $j$;
	\item [d)] If the system recommended item $i$ for user $u_1$, it would be a surprising recommendation;
	\item [e)] If it recommended the same item for user $u_2$, it  would not be as surprising, because user $u_2$ knows $j$, a similar item.
\end{itemize}

\premise{\textit{Surprise must be dynamic}: assuming that surprise depends on beliefs or expectations, it seems reasonable to presume that it changes over time, as the user is constantly being exposed to new experiences.}
This idea can be illustrated through the Scenario 2:

\begin{itemize}
	\item [a)] Suppose items $i$ and $j$ are similar to each other;
	\item [b)] Suppose user $u_1$ has not been exposed to items $i$ and $j$, and all the items known to $u_1$ are very dissimilar from $i$ and $j$;
	\item [c)] Suppose the system recommends item $j$ to user $u_1$;
	\item [d)] After some time has passed item $i$ is recommended to $u_1$;
	\item [e)] Unlike what happened in Scenario 1, recommending item $i$ to $u_1$ is not as surprising, since $u_1$ knows $j$, a similar item.
\end{itemize}
\premise{\textit{Surprise is related to the notion of distance}}: all the metrics reviewed in Section \ref{sec:surprise_metrics} involve the notion of distance; most of them reflect the extent to which a new item resembles the items known to a user.
This is in accordance with the subjective and dynamic views of surprise, since both involve assessing similarity between objects.

In adopting these three ideas as premises for this work, we support the definition of surprise made by \citeauthor{kaminskas2014measuring} \cite{kaminskas2014measuring}, and described in Equation \ref{eq:kam_sidist}.
This definition assumes that the surprise of an item, $S_i(i, E_u)$, is inversely proportional to the degree to which item $i$ is similar to the items known to the user; it adopts a subjective view, since it considers surprise to be a function of the items known to the user.
It also accounts for changes in the surprise of an unobserved item, as the growth of the set of known items.

The remainder of this section has two objectives.
First, to devise a theoretical model that can be used to estimate the total amount of surprise a system can offer to an arbitrary user (Sections \ref{sec:surprise_finite} to \ref{sec:sp_approx}).
Second, to employ the theoretical model to estimate the maximum amount of surprise a system can embed in a recommendation list of arbitrary length (Section \ref{sec:sp_truncated}).
The theoretical model described next has the following settings:
\begin{itemize}
	\item [1)]\premise{Initial condition: each user has rated at least one item;}
	\item [2)]\premise{Interaction: the system produces a recommendation list $L$ to $u$ that only contains one item, which is promptly consumed;}
	\item [3)] \premise{The repository of the system has a finite number of items;}
	\item [4)] \premise{The repository of the system remains stable (no new items are introduced), and this after a finite number of interactions, all the users will have been exposed to all of the items.}
\end{itemize}

\subsection{Surprise is a finite resource}
\label{sec:surprise_finite}
\premise{At any given time, a recommender system has a finite number of items in its repository.}
On the basis of this premise we argue that surprise is a finite resource in this kind of system.
Let $I$ represent the set of items in the repository of the system.
Suppose user $u$ has been exposed to all but one item in the repository, namely item $i$.
Let $N_u = \{i\}$ represent the set of items unknown to $u$, and $E_u = I \, \backslash \, N_u$ the set of items to which user $u$ has been exposed.
Thus, the total amount of surprise the system can offer to $u$ is given by $S_i(i, E_u)$.

This scenario can be modified to allow for $|N_u| > 1$: suppose that user $u$ has been exposed to all but two items, namely $i$ and $j$. Then, $N_u = \{i, j\}$ and $E_u = I \, \backslash \, N_u$.
Suppose that the system recommends items $i$ and $j$ in this order.
Then the total amount of surprise the system can offer to $u$ is $S_i(i, E_u) + S_i(j, E_u \cup \{i\})$. The last term accounts for the fact that item $i$ was known to user $u$ when item $j$ was recommended.
\improvex{LLA: comma between subject and predicate}{It should be noted that the order in which items are recommended may produce a different amount of surprise.}

\subsection{The surprise of a sequence}
Building on the previous scenario, suppose that user $u$ has been exposed to all but $m$ items, namely $N_u = \{i_1, i_2, \ldots, i_m\}$. 
Suppose that the system recommends these $m$ items in a specific order, represented by the sequence $seq = (i_1, i_2, \ldots, i_m)$.
\improvex{LLA: comma between subject and predicate}{Then the surprise of such a sequence of recommendations to the user $u$ can be generalised by the following predicate (surprise of a sequence)}:
\begin{equation}
\label{eq:ss}
S_s(seq, E_u)= S_i(h, E_u) + S_s(t, E_u \cup \{h\}),
\end{equation}
where $h$ represents the head of the sequence $seq$, namely $seq_1$, and $t$ its remaining items, ($seq_2, \ldots, seq_m$).
Since $S_s$ is a recursive predicate, let $S_s(seq, E_u) = 0$ when $|seq| = 0$.

\subsection{The potential surprise}
As stated earlier, the amount of surprise a system can potentially offer to a user $u$, is finite and depends on the sequence in which items are ordered.
\improvex{LLA}{Thus, the maximum potential surprise a system is able to offer to the user $u$ must correspond to the surprise obtained by a specific ordering of the elements of $N_u$}:
\begin{equation}
\label{eq:spmax}
S_{pmax}(N_u, E_u) = \max\limits_{seq \in permut(N_u)} S_s(seq, E_u),
\end{equation}
where $permut(N_u)$ is the set of permutations of items in $N_u$.
The maximum potential surprise, $S_{pmax}$, claims that there are some permutations of the items in $N_u$ that maximise the surprise for the user $u$. This amount of surprise can be interpreted as the ``stock of surprise'' a system can offer to user $u$.
Following the same principle, the minimum potential surprise, $S_{pmin}$, is the permutation of items that minimises the potential surprise for that user:
\begin{equation}
\label{eq:spmin}
S_{pmin}(N_u, E_u) = \min\limits_{seq \in permut(N_u)} S_s(seq, E_u).
\end{equation}

\subsection{The normalised surprise of a sequence}
Once the maximum and the minimum amount of potential surprise a system can offer to a user have been defined, these limits can be used to create a scale that allows the surprise of any sequence comprising all items in $N_u$ to be measured:
\begin{equation}
\label{eq:ssn}
S_{sn}(seq, E_u) = \frac{\hfill S_s(seq, E_u) \hfill - S_{pmin}(seq, E_u)}{S_{pmax}(seq, E_u) \hfill - S_{pmin}(seq, E_u)}.
\end{equation}
The normalised surprise of a sequence, $S_{sn}$, results in a score within the interval $[0, 1]$. If $S_{sn}(seq, E_u) = 1$, then $seq$ is a permutation of items in $N_u$ that maximises the surprise for the user $u$. On the other hand, if $S_{sn}(seq, E_u) = 0$, then it offers the minimum amount of surprise to the user.

\subsection{Computational costs and approximations}
\label{sec:sp_approx}
Real recommender systems have a huge number of items that are unknown to any given user.
Since calculating the $S_{pmax}$ requires evaluating surprise in $|N_u|!$ permutations, its exact computation is not feasible.
However, by applying the optimisation theory to combinatorial problems \cite{johnson1997traveling,mehdi2011parallel}, an approximation to $S_{pmax}$ (Equation \ref{eq:spmax}) can be computed by means of a greedy estimation strategy:
\begin{equation}
\label{eq:spmax-greedy}
\hat{S}_{pmax}(N_u, E_u)=S_i(i^*, E_u) + \hat{S}_{pmax}(N_u \backslash \{i^*\}, E_u \cup \{i^*\})
\end{equation}
\begin{equation}
\label{eq:mostsurprising}
i^* = \argmax_{i \in N_u} S_i(i, E_u).
\end{equation}
In Equation \ref{eq:mostsurprising}, $i^*$ is the most surprising item in $N_u$ with respect to $E_u$.
The same technique can be used to obtain an approximation for $S_{pmin}$ (Equation \ref{eq:spmin}):
\begin{equation}
\label{eq:spmin-greedy}
\hat{S}_{pmin}(N_u, E_u) = S_i(i_*, E_u)+ \hat{S}_{pmin}(N_u \, \backslash \, \{i_*\}, E_u \cup \{i_*\})
\end{equation}
\begin{equation}
\label{eq:leastsurprising}
i_* = \argmin_{i \in N_u} S_i(i, E_u).
\end{equation}
In Equation \ref{eq:leastsurprising}, $i_*$ is the least surprising item in $N_u$ with regard to $E_u$.
We can now define an approximation to $S_{sn}$ (Equation \ref{eq:ssn}) using the approximations for the maximum and minimum potential surprises (Equations \ref{eq:spmax-greedy} and \ref{eq:spmin-greedy}, respectively):
\begin{equation}
\label{eq:ssn_greedy}
\hat{S}_{sn}(seq, E_u) = \frac{\hfill S_s(seq, E_u) \hfill - \hat{S}_{pmin}(seq, E_u)}{\hat{S}_{pmax}(seq, E_u) \hfill - \hat{S}_{pmin}(seq, E_u)}.
\end{equation}

\subsection{Surprise of a recommendation list}
\label{sec:sp_truncated}
Up to this point, we have focused on estimating the total amount of surprise a recommender system can offer to an arbitrary user.
The approach required finding a sequence consisting of all the items unknown to a user ($|seq| = |N_u|$) that maximise the potential surprise for them.
We now turn to the problem of estimating the maximum amount of surprise the system can embed in a sequence that does not contain all the items unknown to a user.
This sequence is referred to as a truncated sequence, and represented as $seq'$.
Now suppose that $seq'$, consisting of $k < |N_u|$ items, obtains the maximum amount of surprise for user $u$ that can be embedded in a sequence with $k$ items.
Then it should be the case that $\hat{S}_{sn}(seq', E_u) = 1$.
Solving for $seq'$ using Equation \ref{eq:ssn_greedy}, give us:
\begin{equation}
\label{eq:surpresas_seqtrunc}
\nonumber
seq' \in \, \left\{\argmax\limits_{q \, \in \, arrangements(N_u, k)} S_s(q, E_u)\right\},
\end{equation}
where $arrangements(N_u, k)$ is the set of $k$-arrangements of items in $N_u$.
An approximation for $seq'$ can be obtained by a greedy strategy.

We recognise that the assumption that $seq'$ can represent a recommendation list $L$ may be subject to criticism, since there are \improvex{too strong?}{important discrepancies} between the settings assumed by the theoretical model and the real user interactions with the the system:

\begin{itemize}
	\item A user may fail to notice a recommendation list $L$, so:\\
	 $userSurprise(L, E_u) = 0 < \hat{S}_{pmax}(L, E_u)$;
	\item A user may not promptly consume all the items in $L$, so:\\
	 $userSurprise(L, E_u) < \hat{S}_{pmax}(L, E_u)$;
	\item A user may not consume an item in the order that it is ranked in the list, so: $userSurprise(L, E_u) < \hat{S}_{pmax}(L, E_u)$.
\end{itemize}
However, in our view, even in such cases, the theoretical model can still be used to provide an upper bound estimation of the surprise experienced by the user.

\subsection{Adapted evaluation method}
Since $\hat{S}_{sn}$ estimates the normalised surprise of a recommendation list, we need a method to assess the performance of a recommender system with regard to this metric.
We adopt the approach employed by \citeauthor{kaminskas2014measuring} \cite{kaminskas2014measuring} and adapt the one plus random off-line evaluation method for this assessment, as described in Algorithm \ref{alg:one-plus-random-ssn}.
It has five parameters, namely a sample of users (\texttt{U}), the set of items in the repository (\texttt{I}), user ratings (\texttt{Ratings}), the length of a recommendation list (\texttt{topN}), and a meta model that, given the set of items known to a user, induces a model that computes a score for an arbitrary, unknown item (\texttt{metamodel}).
As will be described in Section \ref{sec:exp2}, this meta model plays a key role in evaluating predictions supported by the theoretical model.

In line 3, $E_u$ is assigned to the set of items known to the user $u$, and in line 4, $N_u$ is assigned to the set of items unknown to them.
In line 6, $L_1$ is a list with 1,000 unknown items, and each of these items is mapped to a tuple $(i, score)$ and the resulting list is assigned to $L_2$ (line 7).
The \textit{score} is produced by the model induced in line 5.
In lines 8 and 9, tuples in $L_2$ are sorted in descending order and the items that rank in the first \texttt{topN} positions are assigned to $seq'$.
Finally, in line 10, the approximate normalised surprise of $seq'$ is computed and accumulated.
The algorithm returns the average amount of normalised surprise obtained from the user sample \texttt{U}.
\begin{algorithm}[htbp]
	\SetKwFunction{random}{random}
	\SetKwFunction{map}{map}
	\SetKwFunction{sort}{sort}
	\SetKwData{U}{U}
	\SetKwData{I}{I}
	\SetKwData{Ratings}{Ratings}
	\SetKwData{topN}{topN}
	\SetKwData{model}{metamodel}
	\SetKwInOut{Input}{inputs}
	\SetKwInOut{Output}{output}	

	\Input{\U, \I, \Ratings, \topN, \model}
	\Output{Estimated system surprise}
	$acc\leftarrow 0$\;
	\ForEach{$u\in $ \U}{
		$E_u \gets \{e.itemID \in \Ratings \, | \,  e.userID = u\}$\;
		$N_u\gets$ \I $\backslash \, E_u$\;
		$\Theta_u \gets \model\,(E_u)$\;
		$L_1 \gets \random\,(N_u, 1000)$\;
		$L_2 \gets \map\,(i \in L_1, \, (i, \,score \gets \Theta_u(i)) \,)$\;
		$L_3 \gets \sort\,(L_2, key = -score)$\;
		$seq' \gets \map\,( \, (i, score) \in L_3[1:\topN], \, i)$\;
		$acc \gets acc + \hat{S}_{sn}(seq', E_u)$\;
	}
	\KwRet $acc \, / \, |U|$
	\caption{Off-line evaluation method to estimate $\hat{S}_{sn}$}
	\label{alg:one-plus-random-ssn}
\end{algorithm}

\section{Experiments and Results}
\label{sec:experiments}
Two experiments were conducted: the first aims to assess the quality of the greedy approximations for maximum and minimum potential surprise; the purpose of the second is to validate predictions of the theoretical model through different choices of recommender algorithm, data representation, and distance function.

\subsection{Evaluating the approximation strategy}
\label{sec:exp1}

\textbf{Method}: a synthetic dataset was employed to assess the differences between the exact computations of maximum and minimum potential surprises and their greedy approximations.\\\\
\textbf{Dataset}: the dataset contains a single user and eleven items, labelled $\{i_1, \ldots, i_{11}\}$.
The user was exposed to one item ($i_1$). 
The items are represented as vectors in $\mathbb{R}^2$, and arbitrarily distributed.\\\\
\textbf{Procedure}: the degree of surprise was measured for all the permutations of the set of unknown items ($N_u = \{i_2, \ldots, i_{11}\}$, $|N_u|! = 10!$), according to $S_s(seq, E_u)$ in Equation \ref{eq:ss}.
The maximum and minimum surprise measurements obtained were recorded.
The greedy approximations for $S_{pmax}$ and $S_{pmin}$, defined in Equations \ref{eq:spmax-greedy} and \ref{eq:spmin-greedy} respectively, were computed and results recorded.\\\\
\textbf{Variations}: since the surprise of a sequence, $S_s(seq, E_u)$, uses a distance function, the procedure was repeated using four different functions: non-normalised Euclidean distance and cosine distance (geometric intuition), Jaccard distance (combinatorial intuition), and Jensen-Shannon divergence (informational intuition).
The Jaccard distance and Jensen-Shannon divergence to vectors in $\mathbb{R}^n$ were applied as described in \cite{jurafsky2014speech}.\\\\
\textbf{Results}: Table \ref{tab:exp1} shows the values that were calculated using both exact and approximate predicates.
Except for the case of $\hat{S}_{pmax}$ where non-normalised Euclidean distance was used, no substantial difference between the exact and approximate computations was obtained.
That deviation occurred because $\hat{S}_{pmax}$ underestimated $S_{pmax}$.
Since underestimating $S_{pmax}$ or overestimating $S_{pmin}$ could lead $\hat{S}_{sn}$ to achieve a score outside the interval $[0, 1]$, in practice it seems reasonable to clip its value if it falls outside this interval, and this solution was adopted in Experiment 2.
Although these results do not support the general claim that a greedy algorithm will always achieve approximations as good as those obtained, it suggests that the local approximation approach is feasible in real settings.

\begin{table}[H]
	\centering
	\caption{Greedy approximations for potential surprise.}
	\begin{tabular}{ccccc}
		\hline
		Distance&$S_{pmax}$&$\hat{S}_{pmax}$&$S_{pmin}$&$\hat{S}_{pmin}$\\
		\hline
		Euclidean&\textbf{37.684}&\textbf{36.948}&23.935&23.935\\
		cosine&1.367&1.367&0.277&0.277\\
		Jaccard&3.784&3.784&2.552&2.552\\
		Jensen-Shannon&2.089&2.089&0.494&0.494\\
		\hline
	\end{tabular}
	\label{tab:exp1}
\end{table}

\subsection{Evaluating the theoretical model}
\label{sec:exp2}
The theoretical model supports the following predictions:
\begin{itemize}
	\item If a recommender system embeds the maximum amount of surprise in each recommendation it produces, then its evaluation should achieve the maximum score in the potential surprise scale (mean $S_{sn} = 1$).
	\item If a system embeds the minimum amount of surprise in each recommendation, then its evaluation should obtain the minimum value of the scale (mean $S_{sn} = 0$).
	\item Any recommender whose objective is neither to maximise nor minimise surprise, will achieve an intermediate value within the scale ($0 \le \mbox{mean }S_{sn} \le 1$).
\end{itemize}
These predictions should be confirmed regardless of the choices of data representation and distance function.\\\\
\textbf{Method}: a controlled environment was created to enable this experiment to be carried out.
This environment submits data from the MovieLens-1M dataset to a process that produces a time series consisting of measurements for surprise, according to $\hat{S}_{sn}$ (Equation \ref{eq:ssn_greedy}).\\\\
\textbf{Dataset}: the MovieLens-1M dataset was employed \cite{harper2015movielens}. It contains 3,883 items, 6,040 users and just over 1 million ratings. It was enhanced by short movie descriptions collected from the online MovieLens system in September 2017.
Items whose short description was not available or was not written in English were rejected.\\\\
\textbf{Process}: the dataset is submitted to a process comprising three stages: preprocessing, segmentation and measurement.
In the preprocessing stage, a distance matrix is computed for each pair of items in the dataset, by following the parameters specified in each variation outlined in the next paragraph.
During the segmentation stage, the ratings are ordered by timestamp and aggregated into timeframes that include 1,500 ratings each.
Consecutive timeframes that contain ratings from at least 30 common users are marked as eligible for measurement if there is at least one 5-star rating for each user.
These criteria enable us to: 1) control the number of measurement samples; 2) control variation among the samples, since the users that are randomly allocated to a timeframe $T_i$, will be preferably allocated to $T_{i+1}$; 3) satisfy the conditions required by the original one plus random method \cite{cremonesi2008evaluation, bellogin2011precision}, by allowing us to obtain recall evaluations for each sample, for future work.
Finally, during the measurement stage, Algorithm \ref{alg:one-plus-random-ssn} is applied to each eligible interval.
An interval is a sequence of consecutive timeframes, starting from the first timeframe and stretching to an eligible timeframe.
The measurements are sequenced and recorded as a time series.\\\\
\textbf{Variations}: The process was repeated using different choices of recommender algorithm, data representation, and distance function.

\textit{Recommender algorithms}: three algorithms were used: a) the traditional item-kNN (with $k = 50$), which scores items according to the weighted average rating attributed to the $k$ most similar items known to a given user; b) an algorithm that scores items according to their degree of surprise (MSI - Most Surprising Item), which promotes the generation of surprising recommendation lists; and c) an algorithm that scores items according to its familiarity (LSI - Least Surprising Items), which promotes non-surprising recommendation lists.
Since Algorithm \ref{alg:one-plus-random-ssn} is used, a set with 1,000 randomly-sampled unknown items is submitted to each algorithm that, in its turn, attributes a score to each item.
Then the top-ranking items ($\mbox{\texttt{topN}}=10$) are selected as a recommendation list.

\textit{Data representation}: four models were used: Models C and P are semantic models, Model U is a user-item, and Model N is NPMI.

Model C is a traditional vector space model of semantics (or count-based VSM) \cite{baroni2014don,turney2010frequency}.
It uses the short description linked to an item to produce its respective semantic vector.
Items for which the description was too short\footnote{Description is too short if it has less than 13 terms after stop words removal; default NLTK stop words for English were employed \cite{loper2002nltk}.} were rejected.
Before computing tf-idf scores \cite{manning2008introduction}, the terms were stemmed by means of the Snowball algorithm \cite{porter2001snowball,loper2002nltk}.

Model P is a distributed vector space model of semantics (or prediction-based VSM) \cite{baroni2014don}.
It uses the short description linked to each item to produce a semantic vector.
Semantic vectors were extracted through an implementation of the Paragraph Vector \cite{mikolov2013distributed, rehurek_lrec}, which does not require stop words removal or stemming.

Model U is a user-item model \cite{ning2015comprehensive}.
Each item $i$ is represented as a vector $v_i$ of a length equal to the number of users in the system repository. 
Each component $v_{ij}$ represents a) the rating the user $u_j$ has attributed to item $i$, or b) zero if the item was not rated by $u_j$.
The items without any rating were rejected.

Model N is a NPMI score model \cite{kaminskas2014measuring}.
The model consists of two probability distributions, $\mathbb{P}(i)$ and $\mathbb{P}(i, j)$: the first is estimated as the proportion of users who have been exposed to each item $i$, and the latter as the proportion of users who have been exposed to both items $i \mbox{ and } j$.
The distance between $i \mbox{ and } j$ is calculated by means of the NPMI score \cite{bouma2009normalized} (which measures similarity), and then inverted and rescaled so that the image $[-1, 1]$ is mapped to $[0, 1]$.

\begin{table*}[!t]
	\caption{Median, mean and standard deviation of $\hat{S}_{sn}$ over MSI, kNN, and LSI algorithms.}
	\begin{tabular}{cc|ccc|ccc|cccc}
		\hline
		\multicolumn{2}{c}{Variations}&\multicolumn{3}{c}{MSI}&\multicolumn{3}{c}{kNN}&\multicolumn{3}{c}{LSI}\\
		\hline
		Model&Distance&Median&Mean&St.Dev.&Median&Mean&St.Dev.&Median&Mean&St.Dev.\\
		\hline
C&Euclidean&0.912&0.910&0.031&0.448&0.443&0.087&0.023&0.024&0.010\\
C&cosine&0.985&0.980&0.019&0.742&0.740&0.077&0.211&0.219&0.093\\
C&Jaccard&0.969&0.964&0.025&0.704&0.697&0.084&0.172&0.193&0.102\\
C&Jensen-Shannon&0.984&0.975&0.031&0.626&0.615&0.088&0.081&0.097&0.074\\
C&Aitchison&0.979&0.978&0.015&0.512&0.510&0.088&0.037&0.040&0.018\\
\hline
P&Euclidean&0.985&0.984&0.011&0.615&0.605&0.082&0.096&0.099&0.042\\
P&cosine&0.971&0.951&0.061&0.571&0.566&0.083&0.093&0.096&0.050\\
\hline
U&Euclidean&0.921&0.918&0.032&0.835&0.813&0.102&0.004&0.007&0.018\\
U&cosine&0.983&0.970&0.036&0.618&0.633&0.179&0.037&0.042&0.027\\
U&Jaccard&0.999&0.939&0.097&0.585&0.609&0.203&0.053&0.059&0.038\\
U&Jensen-Shannon&0.953&0.948&0.029&0.603&0.602&0.162&0.081&0.085&0.036\\
U&Aitchison&0.946&0.943&0.025&0.758&0.745&0.098&0.008&0.011&0.015\\
\hline
N&NPMI&0.687&0.678&0.091&0.545&0.535&0.111&0.098&0.111&0.072\\
		\hline
	\end{tabular}
	\label{tab:exp2}
\end{table*}

\textit{Distance functions}: six distance functions were used for exploring different intuitions: Euclidean and cosine distances (geometric), Jaccard distance (combinatorial), Jensen-Shannon divergence and NPMI (informational), and Aitchison distance (statistical) \cite{egozcue2011elements}.
The Jensen-Shannon divergence and Aitchison distance are not defined when one vector has a zero component, so they require smoothed vectors.
When needed, smoothing was applied using Bayesian Multiplicative Treatment (BMT) with Perks prior \cite{egozcue2011elements}.
In addition, since there are premises behind each distance, some of them cannot be applied to vectors from all the representation models.
\premise{The Jaccard distance, Jensen-Shannon divergence, and Aitchison distance require compositional data \cite{egozcue2011elements}, which means that they can only be applied to vectors from Models C and U;} and the NPMI score requires an NPMI score model.\\\\
\textbf{Results}: Table \ref{tab:exp2} shows the median, mean, and standard deviation of values obtained from the time series that were produced by executing the process under different variations.
The mean $\hat{S}_{sn}$ under kNN was within the predicted range, but none of the variations under MSI or LSI achieved their predicted results.
The average values for $\hat{S}_{sn}$ under MSI were consistently higher than those under kNN and nearest to 1, whereas the values for $\hat{S}_{sn}$ under LSI were consistently lower than those under kNN and nearest to 0.
There are two reasons for this discrepancy.
First, Algorithm \ref{alg:one-plus-random-ssn} draws 1,000 items from the set of unknown items ($N_u$), while $\hat{S}_{pmax}$ and $\hat{S}_{pmin}$ selects \texttt{topN} items from $N_u$ without sampling.
\improvex{LLA: how to conjugate miss in this case}{In this situation, the probability that a 1,000 sample will miss one of the \texttt{topN} items selected by $\hat{S}_{pmax}$ (or $\hat{S}_{pmin}$) is over 66\%.}
Since the mean $\hat{S}_{sn}$ in Table \ref{tab:exp2} is calculated over a time series comprising 30 intervals (minimum), each containing 30 users (minimum), it means that the probability of all 900 draws will contain all the \texttt{topN} items is practically nil.
Second, Algorithm \ref{alg:one-plus-random-ssn} does not apply a greedy search when selecting items from the sample of 1,000 items, as $\hat{S}_{pmax}$ and $\hat{S}_{pmin}$ do.
\fix{update after E06s1}{A supplementary experiment was conducted, in which Algorithm \ref{alg:one-plus-random-ssn} was altered so that it could use the $N_u$ without sampling and with a greedy search, and then be applied to the variations with the largest discrepancies, namely Model N with NPMI score under MSI and Model C with cosine distance under LSI.}
The results confirmed the predicted results for both MSI and LSI.

\section{Discussion}
\label{sec:discussion}
\theme{BEGIN (interpretive) - Restate your original research questions or statements, followed by main findings, name the results that support your findings}

The aim of this study was to determine a) if there are limits to how much surprise a recommender system can offer to its users, b) how much surprise it can embed in a recommendation list, and c) how these limits can be used to design a metric that reflects how much room there is for improving surprise in recommendations.
While previous studies focused on designing metrics that explore different intuitions about what makes a surprising recommendation \cite{murakami2007metrics,ge2010beyond,akiyama2010proposal,adamopoulos2011unexpectedness,zhang2012auralist,kaminskas2014measuring} or how to combine different metrics \cite{silveira2017framework}, we explored a novel perspective: surprise as a finite resource in any recommender system, whatever intuition about surprise is adopted.

\theme{MIDDLE (interpretive) - report achievements of the study, cite shoulders you climbed, indicate novelty of your work, note similarities in findings, note differences in findings, explain why certain results were obtained}
As the results suggest, there are limits to surprise, and the proposed metric obtained values consistent with these theoretical bounds.
They were also consistent for a number experimental settings where the choices of data representation and distance function varied.
In fact, one contribution made by this work is that it provides further evidence that such choices have a non-negligible effect.
For example, the coefficient of variation of $\hat{S}_{sn}$ over Models C, P and U under kNN, is 13.6\% with cosine distance, and 29.9\% with Euclidean distance.
In addition, the coefficient of variation under kNN for Model U is 13.8\%, and for Model C is 20.7\%.

As briefly discussed in Section \ref{sec:sp_truncated}, \improvex{too strong?}{we recognise that there are important discrepancies between the theoretical model employed in this study and a real-world setting.}
As argued, the model can still be used to estimate an upper bound to the real surprise experienced by a user.
However, there is another limitation that still needs to be addressed.
The theoretical model assumes \evokepr{that the repository remains stable.
This requirement was necessary to allow for a fixed upper bound of potential surprise.} 
As a means of evaluating the impact of undermining this premise, in Experiment 2, the dataset is segmented in a chronological order so that each interval only includes items with an estimated release date within that interval, thus simulating an evolving repository.
The behaviour of the metrics under this condition, as the results suggest, remained within the limits predicted by the theoretical model.

It should also be noted that a surprise model, like any model, is a simplification of the world.
For example, the definition of surprise adopted for this study (Equation \ref{eq:kam_sidist}) models the user experience as a set of items.
As a result, all the things known to a user are represented as points in $E_u$, and all they know is the subject of movies.
The definition also assumes that a user has no bias when recovering from memory the one item that is most similar to that being recommended, according to some notion of similarity.
Both premises are obviously unrealistic, but in our view, these oversimplified models are necessary and still useful, especially in the absence of realistic computational models of surprise that can feasibly be employed to describe an arbitrary user interacting with a recommender system.

\section{Conclusion}
\label{sec:conclusion}
\theme{END - discuss limitations (to deflate criticisms), possible applications of the findings, call for further studies, concluding statement (rename the main finding)}
This study adopts a new approach to evaluate surprise in recommender systems.
\evokepr{A theoretical model of surprise was designed on the assumption that surprise is a limited resource in any recommender system.}
This model predicts the bounds to how much surprise a recommender system can offer its users, and these bounds were employed to design a surprise metric that can be used to determine how competent a system is at embedding surprise in its recommendations and how much room there is for improvement.

Further work should be carried out explore other datasets and recommendation algorithms, as well as other the choices of local optimisation algorithm.
The decision to approximate the potential surprise by using a greedy algorithm took account of the fact that it is easy to manually check its results. 
However, this algorithm lacks theoretical limits to its precision, as some alternatives have.

Finally, the theoretical model opens up a line of thought that it may be fruitful to pursue.
On the assumption that surprise arises from a lack of information, the concept of maximum potential surprise can be framed as the total amount of information a system can offer to a given user.
Since the order in which the items are recommended to a user can lead to a lower amount of experienced surprise, does this mean that information is lost?
It might be the case that this imbalance between maximum potential surprise and actual user surprise, can establish a close relationship with relevance or other properties of recommender systems.
In a loose analogy with mechanical systems, potential energy is never lost; rather, it can only be transformed into something else.

\bibliographystyle{ACM-Reference-Format}
\bibliography{references}

\end{document}